\documentclass{emulateapj}
\usepackage{natbib}
\usepackage{lscape}

\def\gsim{\;\rlap{\lower 2.5pt
 \hbox{$\sim$}}\raise 1.5pt\hbox{$>$}\;}
\def\lsim{\;\rlap{\lower 2.5pt
   \hbox{$\sim$}}\raise 1.5pt\hbox{$<$}\;}
\def\micron{~$\mu\textrm{m}$ }
\def\micronend{$\mu\textrm{m}$}

\begin{document}

\title{The Physical Scale of the Far--Infrared Emission in the Most Luminous Submillimeter Galaxies}
\author{Joshua D. Younger\altaffilmark{1,2}, Giovanni G. Fazio\altaffilmark{1}, David J. Wilner\altaffilmark{1}, Matthew  L. N. Ashby\altaffilmark{1}, Raymond Blundell\altaffilmark{1}, Mark A. Gurwell\altaffilmark{1}, Jia--Sheng Huang\altaffilmark{1}, Daisuke Iono\altaffilmark{3,4}, Alison B. Peck\altaffilmark{5}, Glen R. Petitpas\altaffilmark{1}, Kimberly S. Scott\altaffilmark{6}, Grant. W. Wilson\altaffilmark{6}, \& Min S. Yun\altaffilmark{6}}
\altaffiltext{1}{Harvard--Smithsonian Center for Astrophysics, 60 Garden Street, 
Cambridge, MA 02138}
\altaffiltext{2}{jyounger@cfa.harvard.edu}
\altaffiltext{3}{Institute of Astronomy, The University of Tokyo, 2--21--1 Osawa Mitaka, Tokyo, 181--0015, Japan}
\altaffiltext{4}{Current Address: National Radio Observatory, NAOJ, Minamimaki, Minamisaku, Nagano, 384-1305, Japan}
\altaffiltext{5}{Joint ALMA Office, El Golf 40, Las Condes, Santiago 7550108, Chile}
\altaffiltext{6}{Astronomy Department, University of Massachusetts, Amherst, MA, 01003}

\begin{abstract}

We present high resolution submillimeter interferometric imaging of two of the brightest high--redshift submillimeter galaxies known: GN20 and AzTEC1 at 0.8 and 0.3 arcsec resolution respectively.  Our  data -- the highest resolution submillimeter imaging of high redshift sources accomplished to date -- was collected in three different array configurations: compact, extended, and very extended.We derive angular sizes of 0.6 and 1.0 arcsec for GN20 and 0.3 and 0.4 arcsec for AzTEC1 from modeling their visibility functions as a Gaussian and elliptical disk respectively.  Because both sources are B--band dropouts, they likely lie within a relatively narrow redshift window around $z\sim 4$, which indicates their angular extent corresponds to physical scales of 4-8 and 1.5-3 kpc respectively for the starburst region.  By way of a series of simple assumptions, we find preliminary evidence that these hyperluminous starbursts -- with star formation rates $>1000$ $M_\odot$ yr$^{-1}$ -- are radiating at or close to their Eddington limit.  Should future high resolution observations indicate that these two objects are typical of a population of high redshift Eddington--limited starbursts, this could have important consequences for models of star formation and feedback in extreme environments.

\end{abstract}


\section{Introduction}
\label{sec:intro}

\begin{deluxetable*}{cccccccc}
\tablewidth{0pt}
\tablecaption{Track Details}
\tablehead{
\colhead{Target} & \colhead{Configuration} & \colhead{$u-v$ Coverage} & \colhead{Beam Size} & Date & \colhead{$<\tau_{\rm 225GHz}>$}  & Obs. Time$^a$ & \colhead{Reference$^b$} \\
\colhead{} & \colhead{} & \colhead{[k$\lambda$]} & \colhead{[arcsec]} & \colhead{[dd.mm.yy]} & \colhead{} & \colhead{[hrs]} & \colhead{}
}
\startdata
GN20 & COM & 15-75 & $2.99\times 2.26$ & 20.02.05, 05.03.05 & 0.04, 0.06 & 10.4 & I06 \\
 & EXT & 40-200 & $0.81\times0.75$ & 10.02.08, 11.02.08 & 0.04, 0.04 & 5.3 & This work \\
\hline
AzTEC1 & COM & 20-75 & $2.69\times 2.19$ & 17.01.07 & 0.05 & 5.6 & Y07 \\
 & EXT & 50-250 & $0.86\times 0.55$ & 16.01.08 & 0.03 & 4.0 & This work \\
& VEX & 60-550 &  $0.25\times 0.35$ & 05.04.08 & 0.03 & 3.4 & This work 
\enddata
\tablenotetext{a}{Total on--source integration time in that configuration.}
\tablenotetext{b}{I06: \citet{iono2006}; Y07: \citet{younger2007}}
\label{tab:tracks}
\end{deluxetable*}

\begin{figure*}
\plottwo{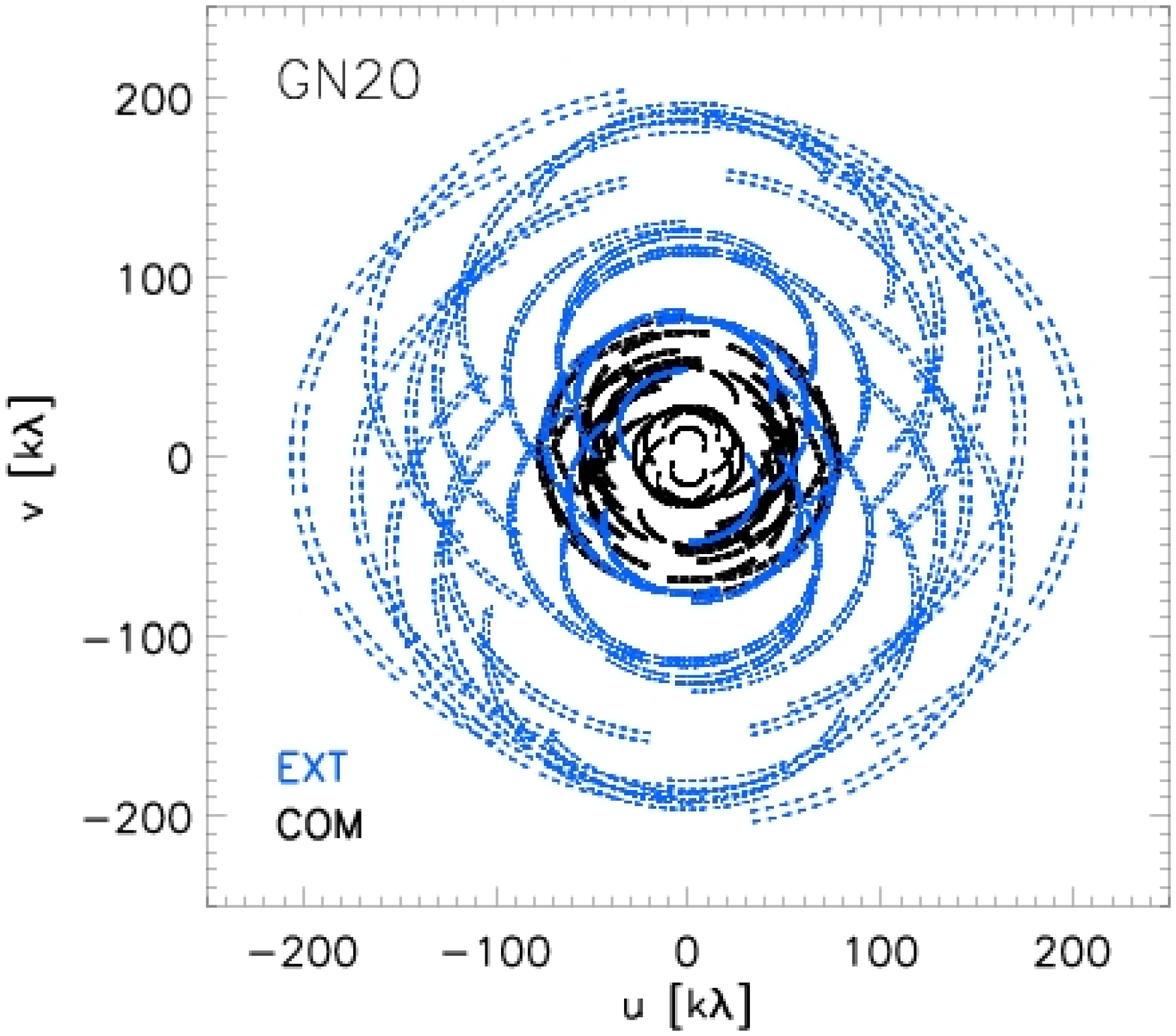}{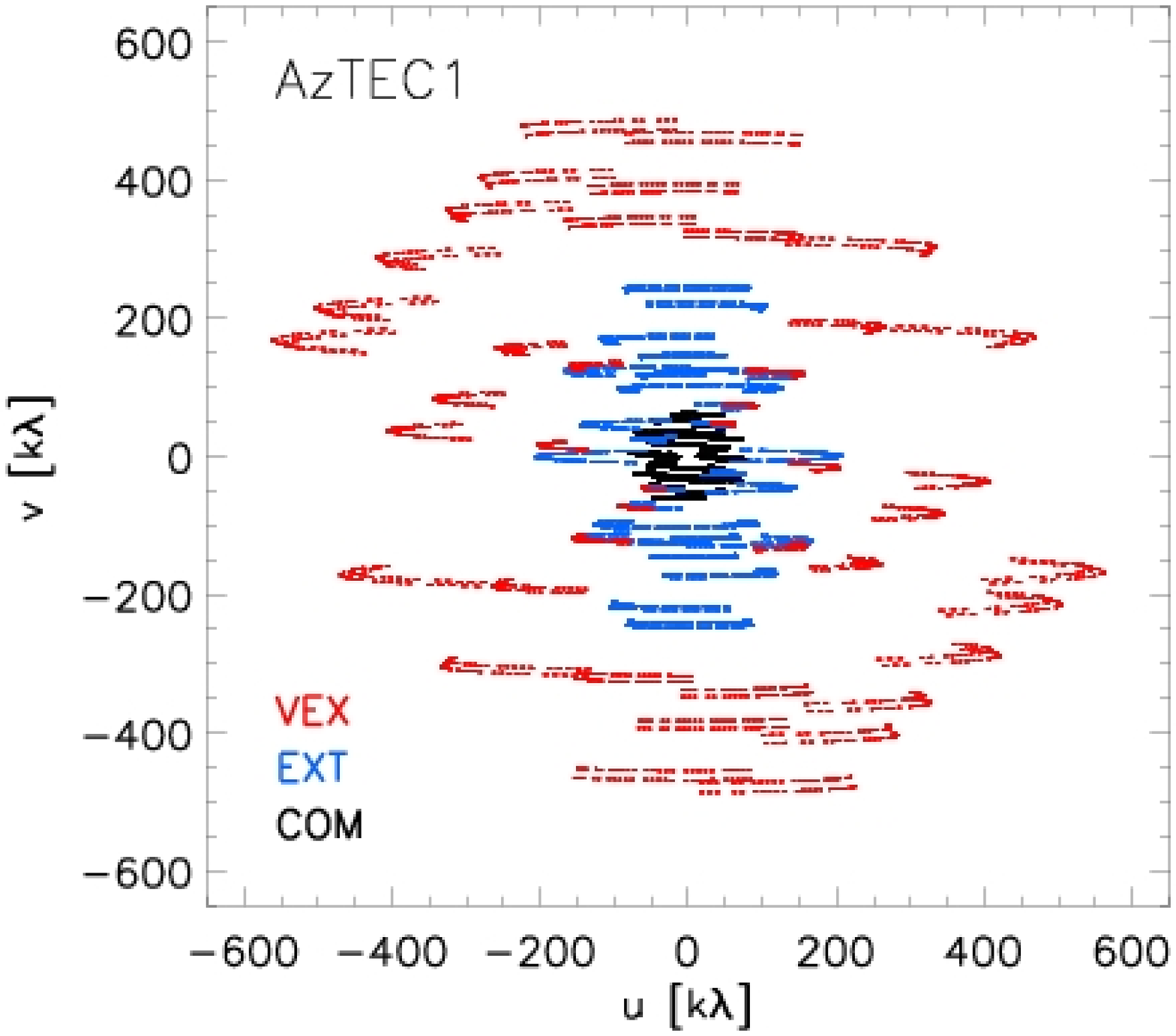}
\caption{The $u-v$ coverage for our high resolution interferometric imaging of GN20 (left) and AzTEC1 (right).  Included are all tracks in three different SMA configurations: compact (COM: black), extended (EXT: blue), and very extended (VEX: red).  For further details, including weather conditions and on-source integration times, see Table~\ref{tab:tracks}.}
\label{fig:uv}
\end{figure*}

Wide area surveys at millimeter \citep[e.g.,][]{greve2004,bertoldi2007,scott2008} and submillimeter \citep[e.g.,][]{smail1997,hughes1998,barger1998,pope2006,coppin2006} wavelengths have revealed a large population of ultra-- and hyperluminous infrared galaxies -- ULIRGs and HyLIRGs -- at high redshift \citep[median $z\sim 2$ for a radio--selected sample][]{chapman2005}.  Since their initial discovery it has become clear that these ``submillimeter galaxies" (SMGs) are likely massive, gas--rich merging systems \citep{frayer1998,frayer1999,chapman2003b,greve2005,tacconi2006,tacconi2008} that represent massive galaxies in formation \citep{scott2002,blain2004}.  Extremely luminous infrared objects take on increasing cosmological importance at $z\gsim 1$ \citep{sanders1996,lefloch2005}, and may dominate cosmic star formation for up to the first half of the lifetime of the universe \citep{blain1999,blain2002}.

Despite significant progress over the past decade, a more complete understanding of SMGs has been hampered in part by the relatively poor resolution of submillimeter cameras ($\sim 10-18$ arcsec FWHM).  In particular, the size scale of the starburst region -- traced by the rest frame far infrared (IR) -- potentially provides important insights into the nature of the engine driving the tremendous luminosity of these systems.  If they are scaled up versions of local ULIRGs, we would expect far--IR emission on scales of $\sim 5-10$ kpc \citep[e.g.,][see also Iono et al. 2008, in preparation]{downes1998,iono2007}.  Hydrodynamic modeling of merger driven nuclear starbursts \citep[e.g.,][]{mihos1994} of the kind thought to drive many SMGs \citep{chapman2003b,tacconi2006,tacconi2008} can be somewhat more compact, a result that could have important physical consequences: Eddington arguments suggest a minimum size scale for such regions \citep{murray2005,thompson2005}.  Unfortunately, at typical SMG redshifts, all of these size scales are far smaller than the typical angular resolution of submillimeter cameras on single--dish instruments.

The first breakthrough came with deep radio continuum surveys, which leveraged the local far--IR/radio correlation \citep[][]{condon1992} in combination with statistical arguments \citep{ivison2002,ivison2007} to associate faint radio counterparts within the submillimeter beam with SMGs.  Higher resolution radio imaging of these sources \citep{chapman2004,biggs2008} found a range of source structures on physical scales of $\sim 1-8$ kpc with a median of 5 kpc.  While promising, this technique assumes a spatially resolved far--IR/radio correlation which is not particularly well understood locally \citep[e.g.,][]{hippelein2003,murphy2006,tabatabaei2007}.  As a consequence, these results are not straightforward to interpret.

This motivates high--resolution imaging of the rest frame far--IR directly, via submillimeter interferometry.  The vast majority of previous work was done at resolutions of $\sim 1-2$ arcsec, and found that the far--IR continuum in SMGs originates at physical scales of $\lsim 4-8$ kpc \citep{neri2003,greve2005,tacconi2006,wang2007,younger2007,younger2008,dannerbauer2008}.  More recently, very high resolution CO imaging by \citet{tacconi2008} showed that gas motions in typical SMGs are disordered on scales of $\sim 1-2$ kpc, suggesting that they are ongoing major mergers.  

In this paper, we present high resolution (beam size $\lsim 1$ arcsec) 890\micron continuum imaging of two of the brightest SMGs known -- GN20 \citep{pope2006,iono2006} and AzTEC1 \citep{younger2007,scott2008} -- with the Submillimeter Array \citep[SMA:][]{ho2004}.  By targeting the brightest -- and therefore likely the most luminous \citep{blain1993} -- objects, we constrain the physical scale of the far--IR in extreme conditions. In addition, since these objects are thought to lie at higher redshift than radio--selected samples \citep{younger2007}, they offer an intriguing probe of the nature of star formation at earlier epochs.  Throughout this work we assume a concordance cosmology with $\Omega_m = 0.3$, $\Omega_\Lambda = 0.7$, and $h=0.7$. 

\section{Observations and Data Reduction}
\label{sec:obs}

\begin{figure*}
\plotone{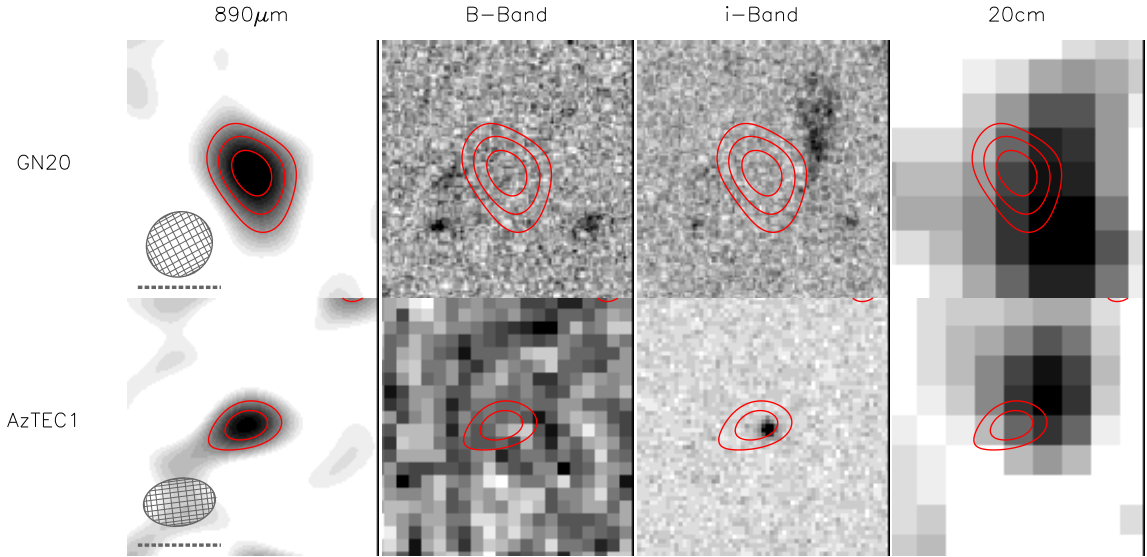}
\caption{Stamp images, 3 arcsec on a side, of GN20 (top) and AzTEC1 (bottom): (from left to right) SMA 890\micron dirty map using EXT configuration data, B--band imaging (HST/ACS for GN20, ground based Subaru for AzTEC1), i--band HST/ACS imaging, and VLA 20cm imaging data.  The contours overlaid on all frames are from the SMA EXT configuration dirty map, in intervals of 3,5,7... times the r.m.s. noise.  For reference, the grey dashed line is 1 arcsec in length and the beam is indicated for each with a hashed grey ellipse: $0.81\times0.75$ and $0.86\times0.55$ arcsec FWHM for GN20 and AzTEC1 respectively.}
\label{fig:stamps}
\end{figure*}

The two targets were GN20 -- the brightest 850\micron source in the Submillimeter Common--User Bolometric Array \citep[SCUBA:][]{holland1999} survey of the Hubble Deep Field North \citep[HDFN: see][]{pope2006}, and AzTEC1 -- the brightest 1.1mm source in the AzTEC \citep{wilson2008} survey of the COSMOS field \citep{scott2008}.  Both were previously detected as single point--sources with flux densities of $F_{890\mu m} = 22.9\pm 2.8$ \citep{iono2006} and $15.6\pm 1.1$ mJy \citep{younger2007} respectively with the SMA in compact configuration (COM).  We have re--observed both these targets with the SMA in extended configuration (EXT), which provides a $\sim 3\times$ improvement in angular resolution over COM, using the same pointing center as the COM tracks.  The EXT tracks -- two for GN20, one for AzTEC1 -- were taken in excellent weather in January and February 2008.  Since AzTEC1 was unresolved in the EXT track (see \S~\ref{sec:results} and Figure~\ref{fig:vis_targets}), we re--observed it in very extended configuration (VEX) in April 2008, which provided a further $\sim 3\times$ improvement in angular resolution.  For details on the tracks, configurations ($u-v$ range, beam size, etc.), and observing conditions, see Table~\ref{tab:tracks}.  

The receiver was tuned to 345 GHz in the USB, and averaged with the LSB for an effective bandwidth of 4 GHz centered at 340 GHz.  For GN20, passband calibration was done using 3C273 and 1921-293, and primary flux calibration was done using Titan.  The target was observed on a 10 minute cycle -- 5 minutes on source, 5 minutes on calibrators -- with two primary gain calibrators: 1048+717 ($\sim 0.3$ Jy; 14 degrees away) and  1153+495 ($\sim 0.3$ Jy; 14 degrees away).  For AzTEC1, passband calibration was done using 3C111 and 3C273, and primary flux calibration was done using Ceres.  As with GN20, the target was observed on a 10 minute cycle with two primary gain calibrators: 1058+015 ($\sim 2$ Jy; 15 degrees away) and 0854+201 ($\sim 2$ Jy; 24 degrees away).  Because Ceres is known to be variable at the $\sim 20-30\%$ level due to rotation \citep{altenhoff1994,redman1998,barrera2005}, we confirm this flux scale by checking that the flux density for 0854+201 derived from this track ($F_{\rm 340 GHz} = 2.37$ Jy) is consistent to that measured one day earlier ($F_{\rm 340 GHz} = 2.29\pm0.12$ Jy) for which Titan was the primary flux calibrator.  

In addition to the two primary targets, we observed a nearby test quasar once every 60 minutes throughout the track to empirically verify the phase transfer and inferred source structure, and estimate the systematic positional uncertainty.  The test quasars for GN20 and AzTEC1 were J1302+578 ($\sim0.1$ Jy; 5.5 degrees away) and J1008+063 ($\sim0.2$ Jy; 5 degrees away) respectively.  Both are included in both the JVAS \citep{patnaik1992,browne1998} and VLBA Calibrator \citep{ma1998,beasley2002} surveys of compact, flat--spectrum radio sources, and have absolute positions known to better than 20 mas. 

For the VEX track on AzTEC1, time dependent gain calibration dervied from
1058+015 left clear, slow, residual phase variations on 0854+201 (and
J1008+063) due to uncertainties in the baseline parameters or other
limitations of the SMA interferometer model. To improve the phase transfer
and prevent decorrelation, an additional gain calibration was performed using
J1008+063, just 4 degrees away from AzTEC1 in declination, since conditions were good enough
to yield sufficient signal-to-noise ($>10\sigma$) in each of the hourly scans
on this source.  This additional step minimized the phase errors owing to
baseline effects in the calibrated visibilities. Remaining phase fluctuations
dominated by the atmospheric effects on short timescales left an
effective seeing sizescale of $\sim0.08$ arcsec in J1008+063 (see \S~\ref{sec:results} and
Figure~\ref{fig:vis_vex} for further discussion).

We also make use of extensive multiwavelength data in both fields. For the HDFN, this includes HST/ACS B--, V--, i--, and z--band optical \citep{giavalisco2004acs},  IRAC 3.6--8.0\micron and MIPS 24\micron \citep{dickinson2003}, and VLA 20cm \citep{biggs2006} imaging data.  For the COSMOS field \citep[see][for an overview]{scoville2007}, this includes Subaru ground based optical \citep{taniguchi2007}, HST/ACS i--band \citep{koekemoer2007}, IRAC 3.6--8\micron and MIPS 24\micron \citep{sanders2007}, and VLA 20cm \citep{schinnerer2007} imaging.

\section{Results}
\label{sec:results}

\begin{deluxetable*}{cccccccccccccc}
\tiny
\tablewidth{0pt}
\tablecaption{Positions and Source Structure}
\tablehead{
\colhead{Name} & \colhead{Config.$^a$} & \colhead{Model} & \colhead{$\alpha$} & \colhead{$\delta$}  &
\colhead{$\Delta\alpha^b$} & \colhead{$\Delta\delta^b$} & 
\colhead{$F_{\rm 890\mu m}$} & \colhead{$\theta^c_{\rm maj}$} & \colhead{$\theta^c_{\rm min}$} & \colhead{$\phi^d$}  \\
\colhead{} & \colhead{} & \colhead{} & \colhead{[J2000]} &\colhead{[J2000]} & \colhead{[arcsec]} & \colhead{[arcsec]} & \colhead{[mJy]} & \colhead{[arcsec]} & \colhead{[arcsec]} & \colhead{[deg]}
}
\startdata
GN20 & C & Point & 12:37:11.920 & +62:22:12.17 & 0.10$^e$ & 0.10$^e$ & $22.9\pm 2.8$ & \ldots & \ldots & \ldots \\
& E & Gaussian & 12:37:11.898 & +62:22:12.14 & 0.06 & 0.09 & $26.9\pm 5.1$  & $0.8\pm 0.2$ & $0.5\pm 0.2$ & 15 \\
& E & Disk & 12:37:11.897 & +62:22:12.16 & 0.06 & 0.09 & $27.3\pm 4.5$ & $1.4\pm 0.3$ & $0.7\pm0.3$ & 25\\
& C+E & Gaussian & 12:37:11.903 & +62:22:12.16 & 0.06 & 0.09 & $23.9\pm2.6$ & $0.8\pm0.2$  & $0.3\pm0.3$ & 35 \\
& C+E & Disk & 12:37:11.901 & +62:22:12.17 & 0.05 & 0.06 & $24.2\pm2.5$ & $1.2\pm0.3$  & $0.8\pm0.3$ & -25 \\
\hline
AzTEC1 & C & Point & 09:59:42.859 & +02:29:38.21 & 0.11 & 0.20 & $15.6\pm 1.1$ & \ldots & \ldots & \ldots \\
& E & Point & 09:59:42.863 & +02:29:38.19 & 0.07 & 0.07 & $13.8\pm2.3$ & \ldots & \ldots & \ldots \\
& C+E & Point & 09:59:42.863 & +02:29:38.19 & 0.05 & 0.06 & $15.1\pm 1.1$ & \ldots & \ldots & \ldots \\
& V$^f$ & Gaussian & 09:59:42.863 & +02:29:38.20 & 0.04 & 0.06 & $15.1\pm1.1$ &  $\sim 0.3$ & $\sim 0.2$ & 40 \\
& V$^f$ & Disk  & 09:59:42.863 & +02:29:38.20 & 0.04 & 0.06 & $15.1\pm1.1$ &  $\sim 0.4$ & $\sim 0.3$ & 25
\enddata
\tablenotetext{a}{Data restricted to this combination of configurations: COM (C), EXT (E), and VEX (V).  See Table~\ref{tab:tracks} and Figure~\ref{fig:uv} for details.}
\tablenotetext{b}{Combined statistical and systematic uncertainty, where the systematic uncertainty is estimated from the position of the test quasar.}
\tablenotetext{c}{$\theta_{\rm maj}$ and $\theta_{\rm min}$ represent the FWHM or diameter of the major and minor axes for the Gaussian and elliptical disk models respectively.}
\tablenotetext{d}{Position angle.}
\tablenotetext{e}{Since there was no test quasar available for this track, these are just the statistical positional uncertainties from \citet{iono2006}.}
\tablenotetext{f}{Because there was some residual phase error due to atmospheric seeing (see \S~\ref{sec:obs} for details), we chose not to combine the V track results with C and E, which did not show these errors.  Furthermore, we fixed the total flux to that inferred from the CE track.  The $\sim$ indicates that these size measurements are large compared with the effective seeing, but are not as robust as those from E and CE on GN20.}
\label{tab:results}
\end{deluxetable*}

\begin{figure*}
\plottwo{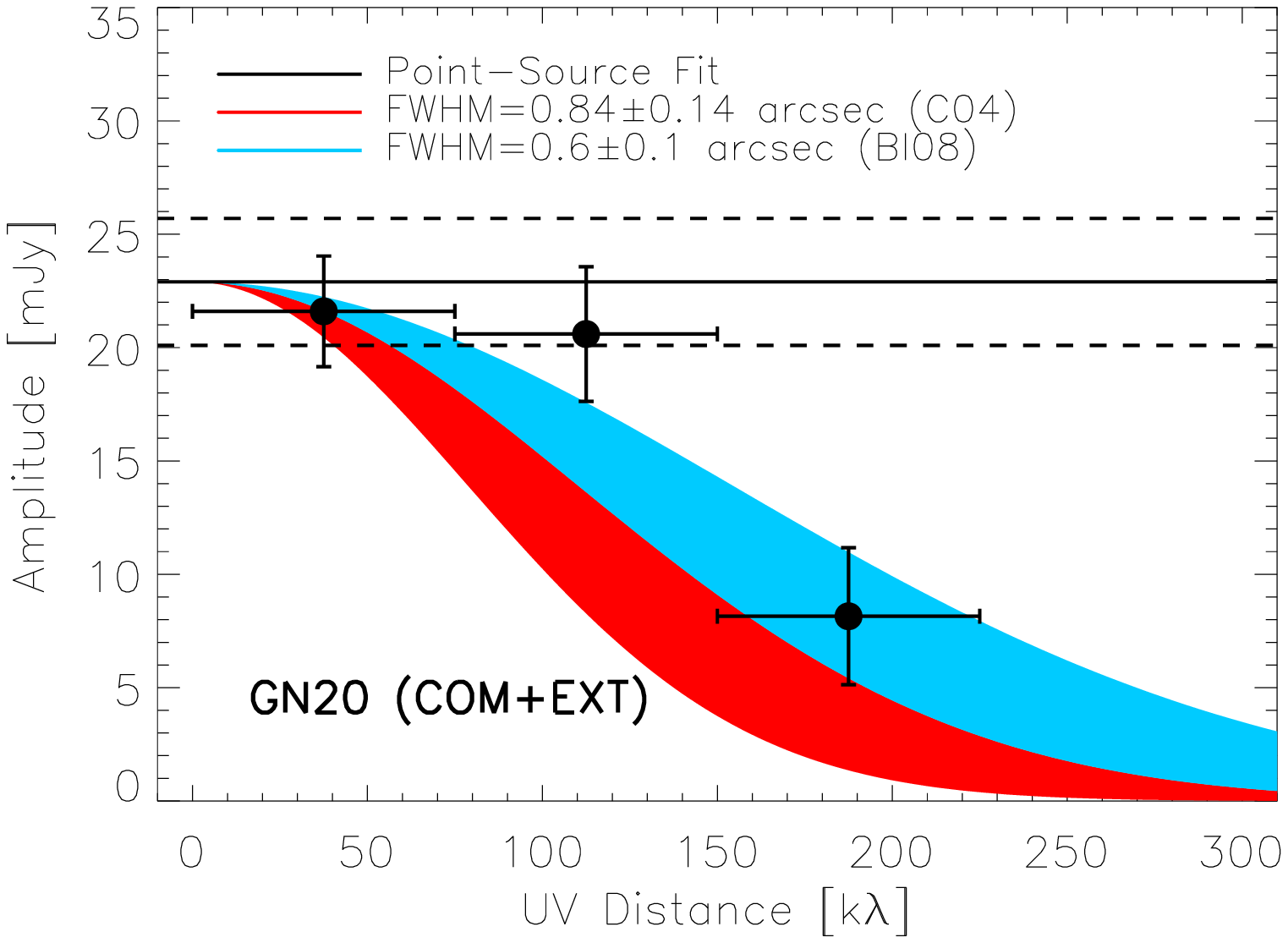}{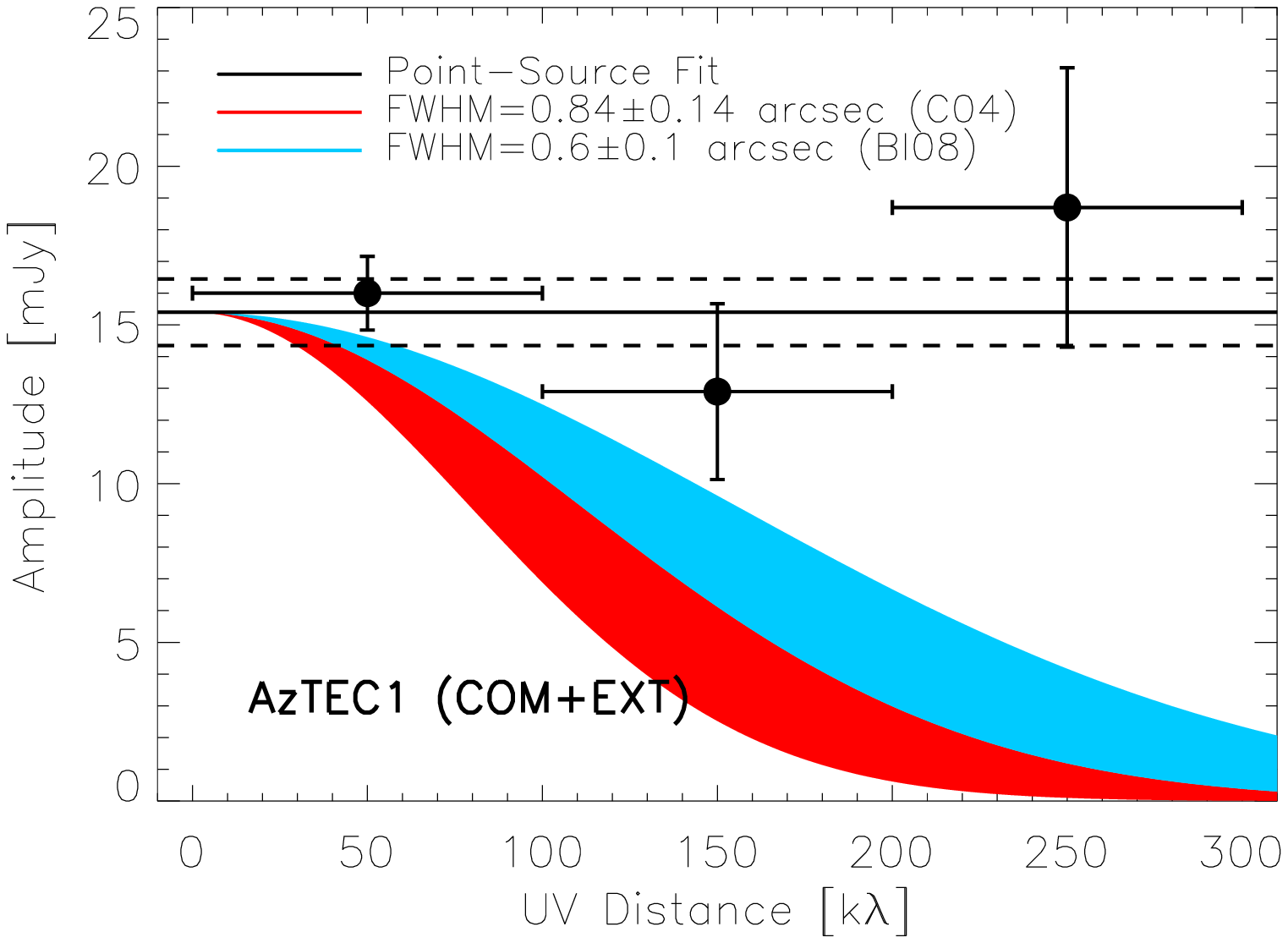}
\plottwo{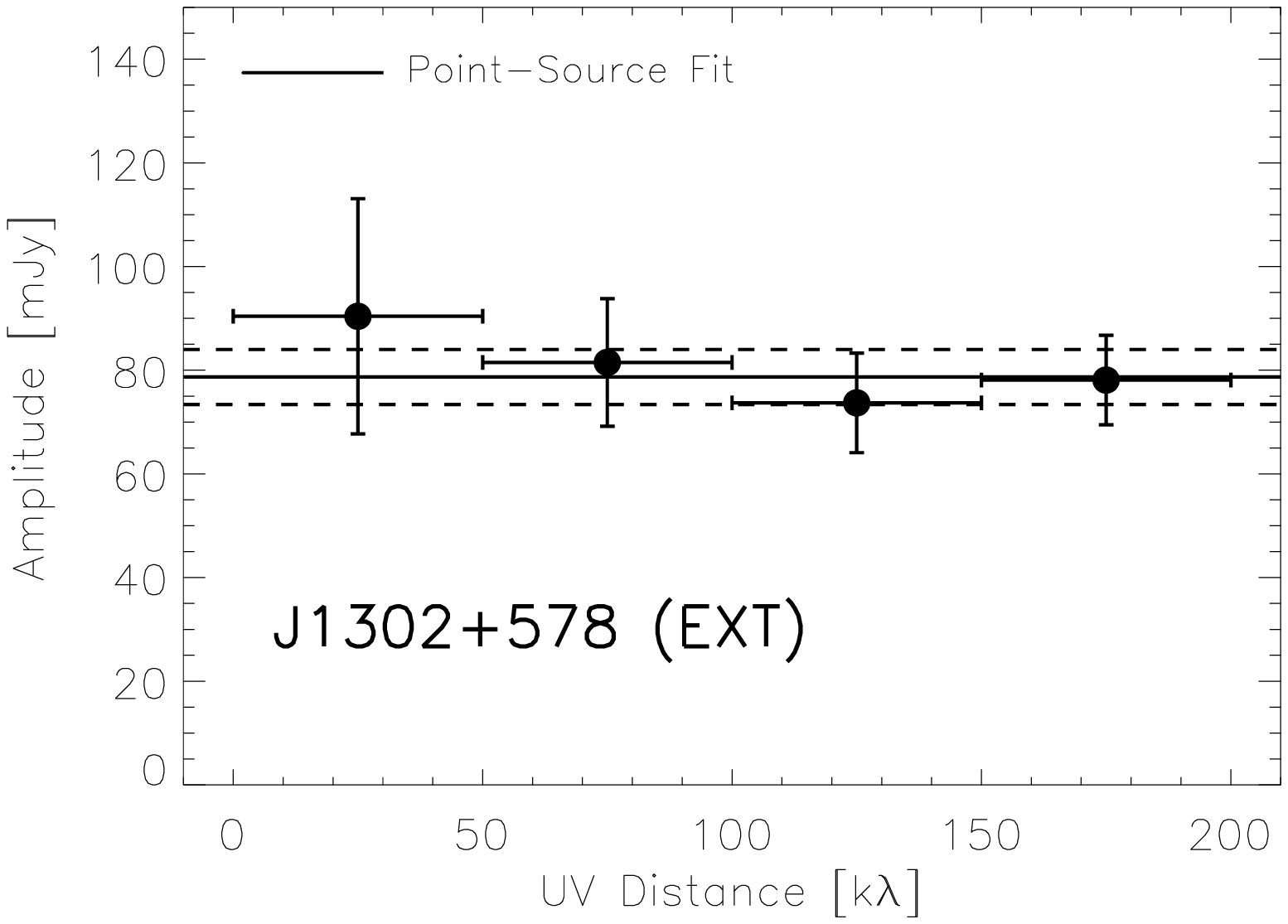}{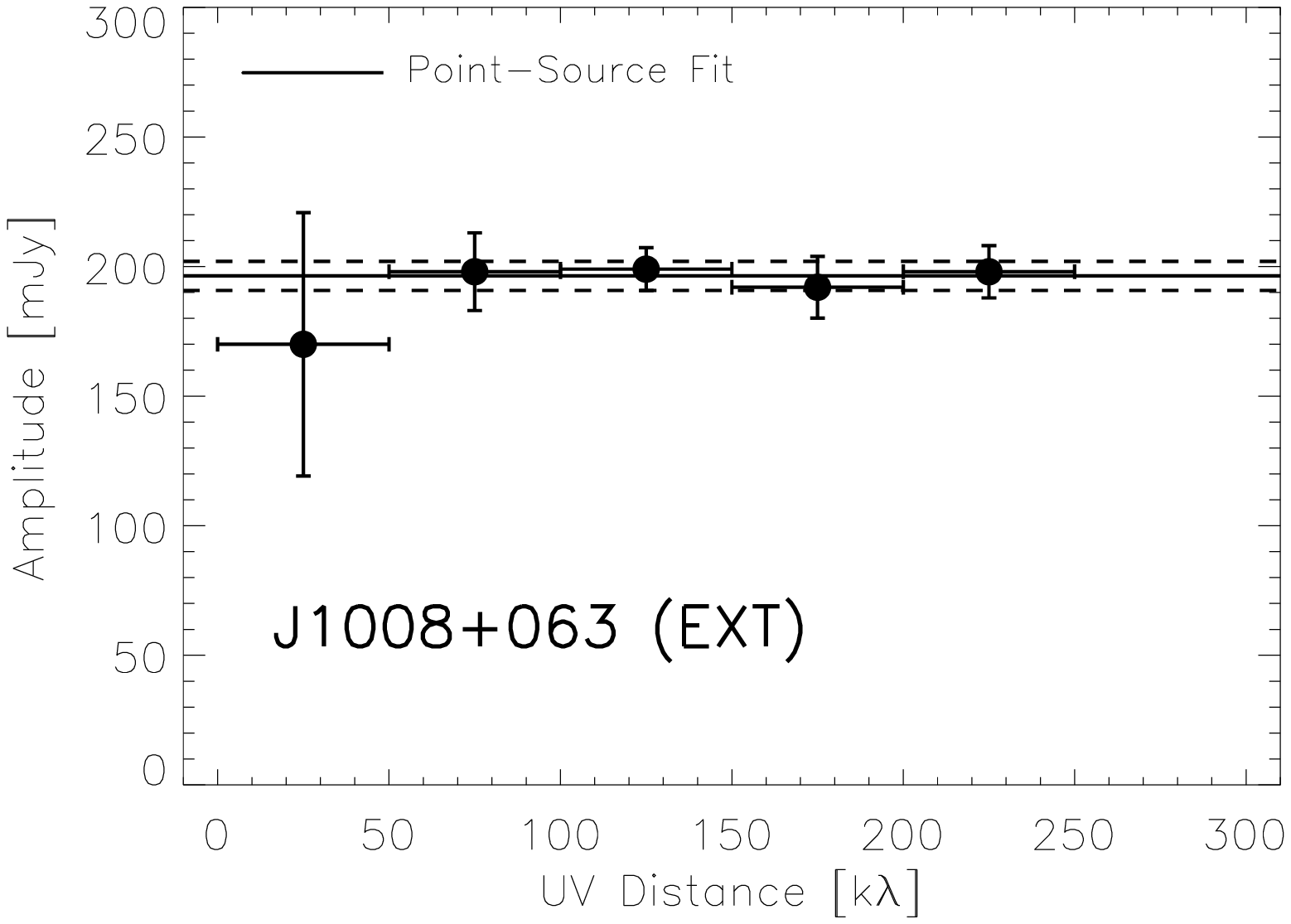}
\caption{{\sc Top:} The real visibility amplitudes as a function of $u-v$ distance -- which shows flux density as a function of decreasing angular scale -- for GN20 (left) and AzTEC1 (right), combining all available data from COM and EXT tracks.  The real part of the $u--v$ data was binned and scalar averaged, with error bars to indicate the dispersion in a given bin.  Also shown are the median radio sizes from \citet[red, C04;][]{chapman2004} and \citet[blue, BI08;][]{biggs2008}.  GN20 shows evidence of being partially resolved on scales of $\sim 200$ k$\lambda$ -- consistent with the median radio sizes of BI08 -- while AzTEC1 is unresolved out to $\sim 250$ k$\lambda$ -- from which we infer a maximum angular scale of $\lsim 0.25$ arsec. The solid line indicates a point--source fit to the visibilities (COM+EXT for AzTEC1, just COM for GN20), with the uncertainty indicated by dashed lines.  {\sc Bottom:} The real visibility amplitudes as a function of $u-v$ distance for test quasars observed during EXT tracks for GN20 (J1302+578; right) and AzTEC1 (J1008+063; right).  This serves as an empirical verification of the phase transfer and source structure.  That both test quasars are unresolved by the SMA, with flat visibility functions, rules out phase errors or seeing effects as artificially imposing structure on our targets.}
\label{fig:vis_targets}
\end{figure*}

\begin{figure*}
\plottwo{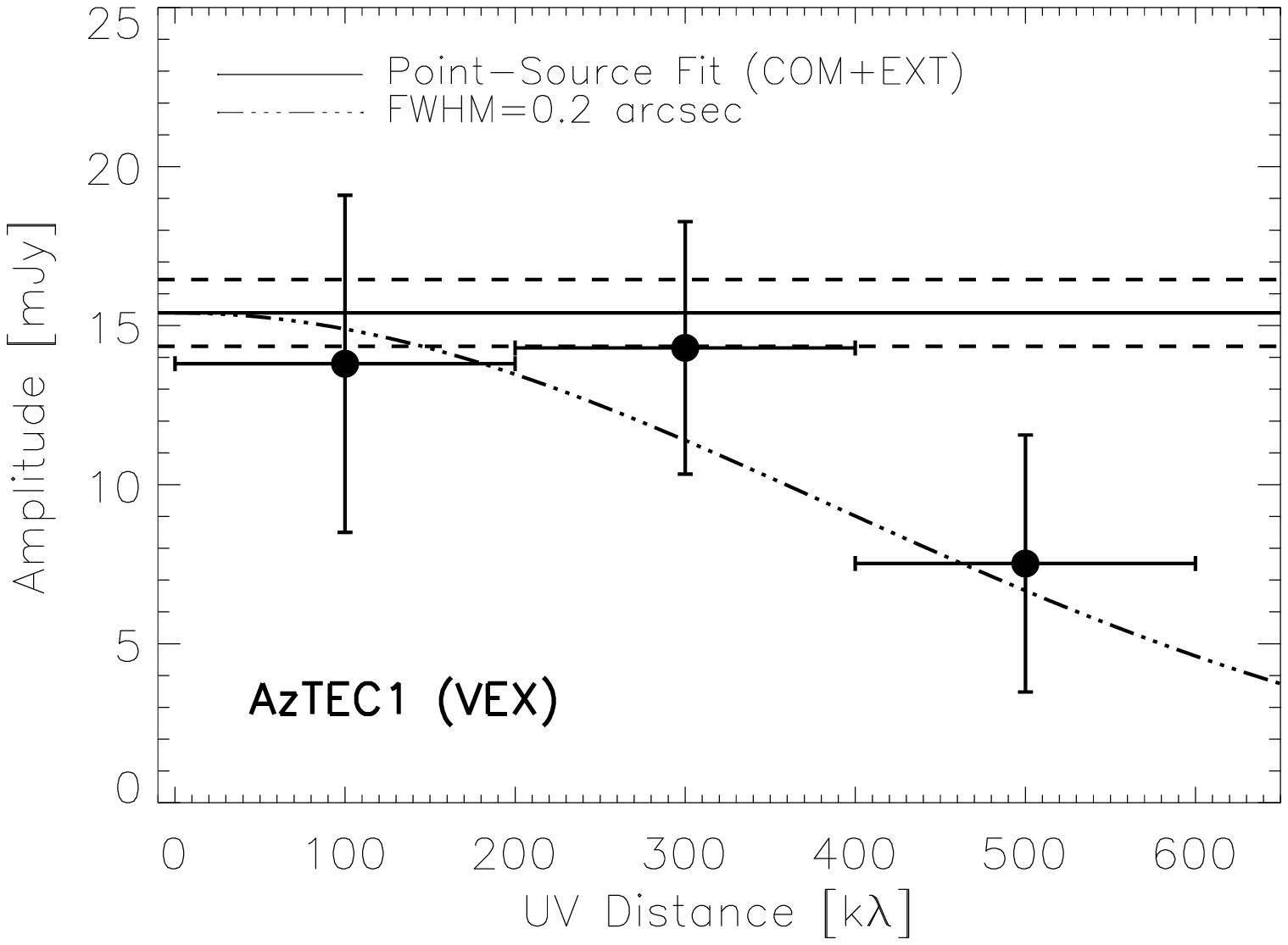}{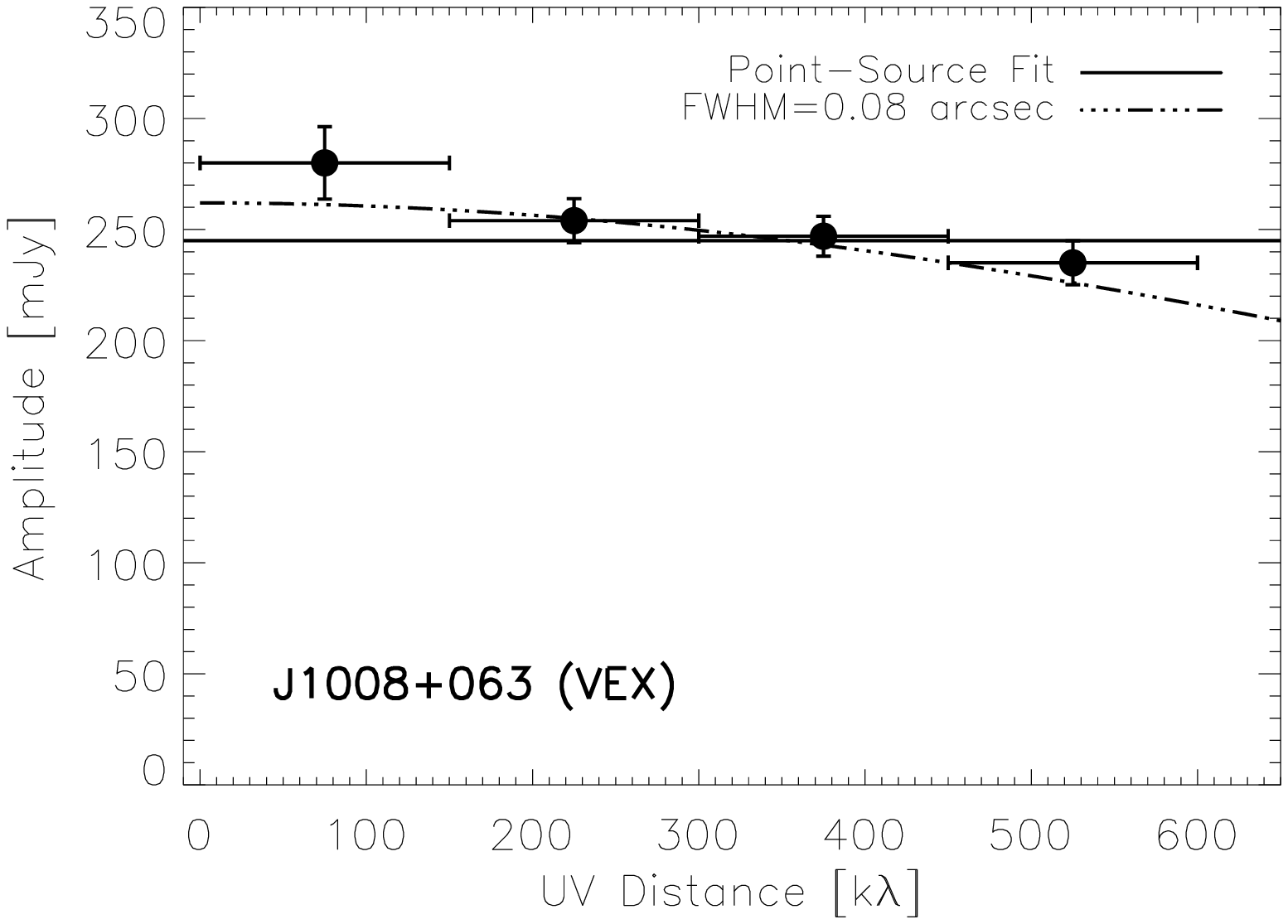}
\caption{The real visibility amplitudes as a function of $u-v$ distance for AzTEC1 (left) and the test quasar J1008+063 (right) using data from the VEX track.  The real part of the $u--v$ data was binned and scalar averaged, with error bars to indicate the dispersion in a given bin.  J1008+063 shows some evidence of partial decorrelation on the the longer baselines, likely a result of residual baseline errors combined with atmospheric effects (see \S~\ref{sec:obs} for a description of the calibration of this track).  The results of a point--source fit are indicated by the solid line, but its visibility function is well--described by a Gaussian of size $\sim 0.08$ arcsec (dot--dot--dashed line), which sets a lower limit on any meaningful source size derived from these data due to this effective seeing.  AzTEC1 shows evidence of being resolved at $\sim 500$ k$\lambda$, or angular scales of $\sim 0.2$ arcsec (dot--dot--dashed line); significantly larger than this seeing size scale and thus a meaningful measurement of the source structure.  Its total flux derived from just this track is consistent with the point--source fit from the COM+EXT tracks (solid line, with dashed lines indicating the uncertainty).}
\label{fig:vis_vex}
\end{figure*}

Both targets were detected at high significance by the SMA in EXT configuration with a $\sim 0.75$ arcsec beam.  The maps, along with overlays on multiwavelength imaging data, are presented in Figure~\ref{fig:stamps}.   Source structure derived from the calibrated visibilities -- which show flux density as a function of decreasing angular scale -- and the empirical verification of phase transfer are summarized in Figures~\ref{fig:vis_targets} and \ref{fig:vis_vex}, and in Table~\ref{tab:results}.

GN20 shows evidence of being partially resolved by the SMA in EXT configuration, with a characteristic angular scale of $\sim 0.5-1.2$ arcsec (see Figure~\ref{fig:vis_targets} and Table~\ref{tab:results}) as inferred from modeling its visibility as both a Gaussian and elliptical disk.  Its submillimeter position is coincident with a bright IRAC 3.6--8\micron and faint MIPS 24\micron \citep[$F_{\rm 24\mu m} \sim 70$ $\mu$Jy;][]{pope2006} source, and roughly so with a radio source ($F_{20cm} = 57\pm 10$ $\mu$Jy; $\sim 0.5$ arcsec away).   The SMA map is also consistent with the radio morphology, which shows some evidence of being resolved along its major axis with a beam size of $1.5\times1.5$ arcsec \citep{biggs2006}.  High--resolution ($0.08$ arcsec PSF) ACS imaging shows that the submillimeter detection is not coincident with the nearby optical ``smudge" .  However, since this source is a B--band dropout, which suggests a redshift range consistent with the observed radio--to--submillimeter \citep[see][]{carilli1999,yun2002} and 24\micronend--to--submillimeter \citep[see][]{wang2007,younger2007} flux density ratios of GN20 (see \S~\ref{sec:discuss}), it is plausible that this object is physically associated with GN20 and represents a region of lower dust opacity.

AzTEC1 is not resolved by the SMA in EXT configuration: its visibility function is flat out to $\sim 250$ k$\lambda$, which suggests a characteristic angular scale of $\lsim 0.5$ arcsec (see Figure~\ref{fig:vis_targets}).  The inferred flux density from the EXT track ($F_{890\mu m}= 13.8\pm2.3$) is furthermore consistent with that from the COM track \citep[$F_{890\mu m}=15.6\pm1.1$ mJy;][]{younger2007}.  The EXT detection is coincident with its compact i--band counterpart in ACS imaging, a faint IRAC 3.6--8\micron source \citep{younger2007}, and roughly so with the radio counterpart ($F_{20cm} = 40\pm13$ $\mu$Jy; $\sim 0.4$ arcsec away) to within the uncertainties -- though Figure~\ref{fig:stamps} appears to show a potentially significant offset between the radio and SMA positions, this is roughly within the total uncertainty in the measurement of their relative position\footnote{The total uncertainty in the offset between the SMA and VLA positions is the combined error from the statistical and systematic uncertainties in the SMA position ($\sigma_{\rm SMA,stat}\sim 0.08$ arcsec and $\sigma_{\rm SMA,sys}\sim 0.05$ arcsec), the systematic uncertainty in the test quasar's absolute position ($\sigma_{\rm VLBA,sys}\lsim 15$ mas), and the statistical and systematic uncertainties in the radio position \citep[$\sigma_{\rm VLA,stat}\approx {\rm HWHM}/({\rm S/N})\sim 0.25$ arcsec; $\sigma_{\rm VLA,sys} \lsim 55$ mas;][]{schinnerer2007}.  This yields a total uncertainty of $\sigma_{\rm tot}\approx 0.4-0.5$ arcsec which is comparable to the observed discrepancy in Figure~\ref{fig:stamps}.}.  It is not detected in the deep COSMOS 24\micron imaging \citep{younger2007}.  The submillimeter size of AzTEC1 is also consistent with its 20cm counterpart, which is compact compared to the $1.5\times1.4$ arcsec VLA beam.  As with GN20, the optical counterpart is a B--band dropout, which suggests a redshift range consistent with the observed radio--to--submillimeter and 24\micronend--to--submillimeter ratios \citep{younger2007}.

The visibility function for AzTEC1 and J1008+063 derived from the VEX track are shown in Figure~\ref{fig:vis_vex}.  Some decorrelation on longer baselines -- likely the result of residual baseline errors in combination with atmospheric effects -- results in artificial structure in the visibility function of J1008+063.  A gaussian fit to this visibility data yields a source size of $(0.09\pm 0.02)\times(0.07\pm 0.02)$ arcsec, which describes the effective seeing size for the track, and thus the minimum source size which is meaningfully probed by these observations.  The visibility function for AzTEC1 shows some marginal evidence of being resolved on scales significantly larger than this lower limit: a Gaussian fit to this visibility data yields a total flux of $16.0\pm 5.0$ mJy -- consistent with the COM, EXT, and COM+EXT fits -- with a size of $(0.30\pm 0.15)\times(0.20\pm 0.10)$ arcsec.  Fixing the total flux to the value derived from the COM+EXT data marginally improves this size measurement to $(0.29\pm 0.13)\times(0.18\pm 0.10)$ arcsec.  While the statistical uncertainty in the position measurement for AzTEC1 also improves to $\sim 0.04$ arcsec in both $\alpha$ and $\delta$, because we calibrate using J1008+063 to remove baseline errors it is exactly at the phase center and therefore does not provide an estimate of the systematic positional uncertainty.  Therefore, we quote the position and flux derived from the COM+EXT tracks in Table~\ref{tab:results}.

\section{Discussion}
\label{sec:discuss}

The robust result of these observations is that the far--IR emission in both GN20 and AzTEC1 is small considering the very high luminosity of these systems, but is clearly extended on $\sim$ kpc scales.  This is suggestive of mergers as the physical mechanism driving the bolometric luminosity of these systems \citep{mihos1994,hopkins2006}.  However in general, and in particular for the case of AzTEC1, this does not require that the far--IR luminosity is contributed only by the starburst.   Indeed, a significant fraction of the far--IR could arise from a dusty torus associated with an active nucleus, which is generated on significantly smaller scales \citep[e.g.,][]{urry1995}.  Ideally one would like significantly improved resolution continuum imaging and resolved gas kinematics via molecular spectroscopy to constrain the structure of these sources and dynamical state of the star--forming gas in detail -- measurements which are beyond the capabilities of current facilities (e.g., SMA, CARMA, PdBI) but in the near term future will likely be accomplished with relative ease by ALMA.  Nevertheless, if we assume that GN20 and AzTEC1 are starburst dominated -- as the typical SMG is thought to be \citep{alexander2005,alexander2008} -- and make a series of admittedly crude but arguably reasonable assumptions about their morphology and kinematics, we find a preliminary indication that they may be radiating close to or at the Eddington limit of their starburst.

It has been suggested that feedback from ongoing star formation sets a physical limit on the minimum size of a star forming region \citep{elmegreen1999,murray2005,thompson2005}.  Owing to the significant opacity of dust to the ultraviolet light produced by young stars, radiation pressure from high luminosity star formation regions can produce strong momentum--driven winds \citep[e.g.,][]{netzer1993,elitzur2001}.  These winds are confined by the gravitational potential, which for an isothermal sphere scales as $\Phi \sim f_g \sigma^2 \log{D}$ -- where $f_g$ is the gas fraction, $\sigma$ is the stellar velocity dispersion, and $D$ is the diameter of the starburst region.  In the optically thin limit -- which is appropriate for optically thick clouds with a small volume filling factor embedded in the diffuse interstellar medium -- and assuming a \citet{salpeter1955} initial mass function (IMF\footnote{Using a \citet{kroupa2001} or \citet{chabrier2003} IMF will tend to lower SFR$_{\rm max}$ by $\sim 40\%$ \citep{kennicutt1998cal,bell2003,bell2005}.}), this leads to a maximum star formation rate (SFR) of:
\begin{equation}
{\rm SFR_{max}} = 900\, \sigma_{400}^2 D_{\rm kpc} \kappa_{100}^{-1}\,\,\, M_{\odot}\,\,{\rm yr^{-1}}
\end{equation}
where $D_{\rm kpc}$ is the characteristic physical scale of the starburst -- measured via either the Gaussian FWHM or disk diameter -- in kpc, $\sigma_{400}$ is the line--of--sight gas velocity dispersion in units of 400 km $s^{-1}$, and $\kappa_{100}$ is the dust opacity in units of 100 cm$^{2}$ g$^{-1}$ \citep{murray2005} .  As an upper limit on SFR$_{\rm max}$, we adopt $\kappa_{100} \approx 1$, but note that many dust models allow for orders of magnitude higher opacity, particularly for the ultraviolet radiation produced by young massive stars during a starburst \cite[e.g.,][]{li2001}.

\begin{figure}
\plotone{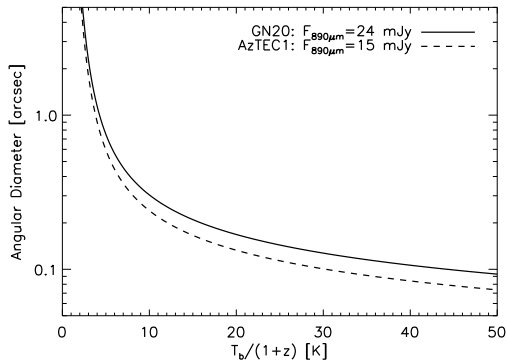}
\caption{The angular diameter required for a rest--frame brightness temperature at cosmological redshift $T_{b}/(1+z)$ for both GN20 (solid line) and AzTEC1 (dashed line).  In the optically thick limit, in which the dense molecular clouds have a high volume filling fraction, the dust temperature $T_d = T_{b}$. For GN20, characteristic angular scale inferred from a Gaussian and disk model yield $T_b \approx 6(1+z)$ and $4(1+z)$ K respectively, which at $z\sim 4$ suggests $T_d \approx 30$ or $T_d \approx 20$.  For AzTEC1, these models yield $T_b \approx 8(1+z)$ and $6(1+z)$ K, which at $z\sim 4$ gives $T_d \approx 40$ and $T_d \approx 30$ K.}
\label{fig:tb}
\end{figure}

Our high resolution observations of AzTEC1 and GN20, which provide a measurement of both the far--IR luminosity and physical scale of the starburst region in two of the most luminous systems known, offer an ideal test--bed for this hypothesis.  Both targets are B--band dropouts (see Figure~\ref{fig:stamps} and \S~\ref{sec:results}), which indicates that they likely lie within a relatively narrow redshift range $3.5\lsim z\lsim4.5$ \citep[see e.g.,][]{steidel1999,giavalisco2004}.  While these colors are possibly consistent with a very dusty $z\sim 2$ source, the lack of a bright MIPS 24\micron counterpart -- arising from redshifted $\sim 8\mu$m PAH emission -- and their observed radio--to--submillimeter flux density ratios -- assuming a far--IR SED similar to Arp 220 -- are consistent with this higher redshift interpretation \citep[][]{younger2007}.  Furthermore template fitting to both the optical and far--IR SEDs of AzTEC1 independently yield a consistent result of $z\sim 4$ (Yun et al., in preparation).   For this range, the angular diameter distance is roughly constant with redshift, so observed angular sizes $\theta$ correspond to physical scales of $\ell \approx 7\, (\theta/{\rm arcsec})$ kpc.  We can also roughly estimate their total far--IR luminosity by assuming an Arp 220 template\footnote{The inferred luminosity is known to be uncertain by a factor of $\sim 2-3$ due to variations in the dust temperature and emissivity \citep[e.g.,][]{blain2003}.  For example, if we assume a Mrk 231 template with significant AGN contribution to the far--IR, the inferred far--IR luminosity and SFR will be $\sim 40-60\%$ lower \citep[e.g.,][]{stevens2005,huang2007}.}, which for $z\gsim 1$ gives $L_{FIR} \approx 2\, (F_{\rm 890\mu m}/{\rm mJy})\,10^{12}\, L_{\odot}$ \citep[see also][]{neri2003} and -- assuming a \citet{salpeter1955} initial mass function (IMF)  -- a star formation rate of ${\rm SFR} \approx 340\, (F_{\rm 890\mu m}/{\rm mJy})$ $M_{\odot}\, {\rm yr}^{-1}$, owing to the strong negative $k$--correction in the submillimeter at high--redshift \citep{blain1993,blain2002}.

CO spectroscopy offers the best route to a measurement of the dynamical state of the star forming gas, including $\sigma_{400}$.  Unfortunately, optical redshifts of the requisite precision are not available for these sources.  Similar observations of other SMGs found typical velocity dispersions of $\sigma_{400} \approx 1$ for somewhat lower luminosity systems \citep[median $S_{\rm 850\mu m} \approx 8-11$ mJy, or $L_{FIR} \approx 2\times 10^{13} L_\odot$;][]{neri2003,greve2005,tacconi2006,tacconi2008}.  This is consistent with the observed connection between SMGs and present day 3--4 $L^\star$ early--type galaxies, in combination with the starburst mass fractions in the remnants of gas--rich mergers expected from simulations and observed in local systems \citep{hopkins2008}.  In what follows, we use these observational constraints and adopt $\sigma_{400} = 1$ with the awareness that this parameter is somewhat uncertain for the more extreme systems we are studying.  

GN20 has a very high far--IR luminosity, with $L_{FIR}({\rm GN20}) \approx 5\times 10^{13}\, L_\odot$ and ${\rm SFR(GN20)} \approx 8000$ $M_\odot$ yr$^{-1}$ on characteristic a physical scale of $\ell_{\rm G}({\rm GN20}) \approx 4$ and $\ell_{\rm D}({\rm GN20}) \approx 7$ kpc for a Gaussian and elliptical disk model respectively.  Adopting $\sigma_{400} = 1$, the  corresponding Eddington limits for each source model are ${\rm SFR_{max,G}} \approx 3600$ and ${\rm SFR_{max,D}} \approx 3600$ $M_\odot$ yr$^{-1}$.  This very luminous SMG is close to or at the Eddington limit for a starburst on those scales.  It is also interesting to note that this size scale is somewhat extended compared to a simple $R\sim L^{1/2}$ -- for a disk geometry -- or $R\sim L^{1/3}$ -- for a spherical geometry -- scaling of the starburst size of local ULIRGs \citep[e.g.,][see also Iono et al. 2008, in preparation]{downes1998,iono2007}.  

AzTEC1 has a far--IR luminosity\footnote{This differs from the luminosity implied by Fig. 3 of \citet{younger2007} because it is derived from the far--IR directly, not from the near--infrared.} of $L_{FIR}({\rm AzTEC1}) \approx 3\times 10^{13}\, L_\odot$ and ${\rm SFR(AzTEC1)} \approx 5000$ $M_\odot$ yr$^{-1}$ on a characteristic physical scales of $\ell_{\rm G}({\rm AzTEC1}) \approx 1.5$ and $\ell_{\rm D}({\rm AzTEC1}) \approx 2.5$ kpc for a Gaussian and elliptical disk model respectively.  Adopting $\sigma_{400} = 1$, this is significantly larger than the Eddington limit for a starburst on this scale, with ${\rm SFR_{max,G}} \approx 1350$ and ${\rm SFR_{max,D}} \approx 2250$ $M_\odot$ yr$^{-1}$.  Increasing the dust opacity only aggravates the situation.  However, some SMGs have been observed with $\sigma_{400} \sim 1.5-2$, which could explain the discrepancy.  However, under the assumption that the far--IR luminosity is dominated by a starburst component, even at high velocity dispersion AzTEC1 is close to its Eddington limit.

\begin{deluxetable}{ccccccccc}
\tablewidth{0pt}
\tablecaption{Two--Component Fitting Results}
\tablehead{
\colhead{Name} & \colhead{$F_{\rm 1, 890\mu m}^a$} & \colhead{$F_{\rm 2, 890\mu m}^b$} & \colhead{$\Delta\theta^c$} \\
& [mJy] & [mJy] & [arcsec] 
}
\startdata
GN20$^d$ & $13.5\pm2.4$ & $9.6\pm2.4$ & $0.6\pm0.13$ \\
AzTEC1$^e$ & $8.9\pm2.2$ & $4.6\pm 2.1$ & $0.3\pm0.11$
\enddata
\tablenotetext{a}{Flux of the first component derived from fitting a two--component point--source model to the calibration visibilities.}
\tablenotetext{a}{Flux of the second component derived from fitting a two--component point--source model to the calibration visibilities.}
\tablenotetext{c}{The separation of the two fitted components, including both the statistical and systematic positional uncertainty (see \S~\ref{sec:obs} for details \citep[see also][]{younger2007}.}
\tablenotetext{d}{Derived from a fit to the COM+EXT data.}
\tablenotetext{d}{Derived from a fit to just the VEX data.}
\label{tab:two_points}
\end{deluxetable}

If the volume filling factor of dense molecular gas is close to unity, and therefore the star forming gas is optically thick, then -- again assuming a \citet{salpeter1955} IMF -- the Eddington limit \citep{murray2005} on the star formation rate is 
\begin{equation}
{\rm SFR_{max,thick}} = \frac{4 f_g c}{G} \sigma^4 \approx 10^5 f_{g,0.5} \sigma_{400}^4 \,\,\, M_{\odot}\,\,{\rm yr^{-1}}
\end{equation}
where $f_{g,0.5}$ is the gas mass fraction in units of 0.5.  This is an order of magnitude higher than the SFR of GN20 and AzTEC1.  However, in the optically thick limit, the dust temperature $T_d$ and brightness temperature $T_b$ -- defined as $I_\nu = B_\nu(T_b)$, where $I_\nu$ is the surface brightness and $B_\nu$ is the Planck function -- are equivalent.  The implied angular diameter $\Theta$ of both GN20 and AzTEC1 for a given brightness temperature at cosmological redshift ($T_b/(1+z)$) is shown in Figure~\ref{fig:tb}. For GN20, characteristic angular scale inferred from a Gaussian and disk model yield $T_b \approx 6(1+z)$ and $4(1+z)$ K respectively, which at $z\sim 4$ suggests $T_d \approx 30$ or $T_d \approx 20$.  For AzTEC1, these models yield $T_b \approx 8(1+z)$ and $6(1+z)$ K, which at $z\sim 4$ gives $T_d \approx 40$ and $T_d \approx 30$ K.  These are all somewhat lower than would be expected from the temperature--luminosity relation at low \citep{dunne2000,klaas2001,yang2007b}, intermediate \citep{yang2007a}, and high redshift \citep{blain2003,chapman2005,kovacs2006}.  However, the brightness temperature represents a lower limit, as the inferred dust temperature will increase as the opacity $\tau_\nu$ decreases  -- i.e., $T_d = T_b(1+z)/(1-e^{-\tau_\nu})$ -- or if the volume filling factor of optically thick clouds is less than unity.  There is evidence that in the cores of local ULIRGs $\tau_{\rm 100\mu m} \lsim 1$ for $\lambda \gsim 100$\micron \citep{solomon1997}, and that the volume filling factor of dense molecular gas is $\approx 30-70\%$ \citep{downes1993}.  By analogy, it is plausible that GN20 and AzTEC1 are intermediate between the optically thick and optically thin regimes.  Future observations at shorter wavelengths -- e.g., at 350\micron with SHARC--II \citep{dowell2003} -- could constrain $T_d$ independently, and thus help determine the appropriate limit.

Furthermore, our $u-v$ coverage does not exclude a multi--component structure for either GN20 or AzTEC1, in particular one with two compact point sources.   The results of a fit to the calibrated visibilities for this model are summarized in Table~\ref{tab:two_points}.   These angular offsets correspond to a physical separation of 4 and 2 kpc for GN20 and AzTEC1 respectively, which are consistent with dual nuclear starbursts in a late stage merger \citep[e.g.][]{mihos1994,hopkins2006}.  Should higher resolution data from either the SMA or ALMA confirm this interpretation, it is possible that each of these components is at or close to its Eddington limit with $\sigma_{400}\gsim 2$.

\section{Conclusion}

We present high resolution interferometric submillimeter imaging of two of the brightest -- and therefore likely most luminous \citep{blain1993} -- high redshift starburst galaxies known -- GN20 and AzTEC1.  The visibility functions for these sources indicate characteristic angular sizes of $\sim 0.5-1.2$ and $\sim 0.2-0.4$  arcsec respectively.  Both are B--band dropout optical sources, which indicates a redshift of $3.5\lsim z\lsim4.5$, and thus these angular size measurement correspond to physical scales of $4-8$ and $1.5-3$ kpc.  Assuming a simple morphology and a dynamical state typical of high--redshift SMGs, we find preliminary evidence that GN20 and AzTEC1 are both close to the limiting luminosity derived via Eddington arguments.  If these two sources are indicative of a large population of hyperluminous starbursts at high redshift, this may have important consequences for models of star formation and feedback in extreme environments. 

\acknowledgements

Thanks to the anonymous referee for their helpful comments, and to T. J. Cox, Philip F. Hopkins, Lars Hernquist, Chris Hayward, Yeuxing Li, and Stephanie Bush for helpful discussions.  We also thank the SMA operators, in particular Zach Gazak and Ryan Howie for their help executing these tracks in excellent conditions.  The Submillimeter Array is a joint project between the Smithsonian Astrophysical Observatory and the Academia Sinica Institute of Astronomy and Astrophysics and is funded by the Smithsonian Institution and the Academia Sinica.  This work is supported in part by a grant from the W.M. Keck Foundation.  

{\em Facilities:} \facility{SMA}, \facility{JCMT}, \facility{Spitzer (IRAC, MIPS)}, \facility{HST (ACS)}, \facility{Subaru (Suprime-Cam)}, \facility{VLA}

\bibliographystyle{apj}
\bibliography{../../smg}

\end{document}